\documentclass[aps,prl,twocolumn,showpacs,superscriptaddress,nofootinbib,floatfix,showkeys,10pt]{revtex4-2}
\usepackage{bm,times,braket,amsfonts,amssymb,stmaryrd,latexsym,amsmath}
\usepackage[usenames,dvipsnames]{color}
\usepackage{epsfig,slashed,hyperref,subfigure,graphicx,orcidlink,multirow}
\allowdisplaybreaks[1]
\definecolor{nicered}{rgb}{0.7,0.1,0.1}
\definecolor{nicegreen}{rgb}{0.1,0.5,0.1}
\definecolor{navyblue}{RGB}{0,110,184}
\hypersetup{colorlinks,citecolor=nicegreen,linkcolor=nicered,urlcolor=navyblue,
pdftitle={Heavy-meson molecules with dynamical pions}}

\newcommand{\eq}{Eq.~\eqref}

\newcommand{\Tcc}{T_{cc}^{+}}
\newcommand{\Xstate}{X(3872)}
\newcommand{\epole}{E_{\mathrm{pole}}}
\newcommand{\NR}{\mathrm{NR}}
\newcommand{\Rpi}{\mathrm{R}}

\newcommand{\dd}{\mathrm{d}}

\graphicspath{{figures/}}

\begin{document}

\title{Heavy-meson molecules with dynamical pions}
\author{Yan-Ke Chen\,\orcidlink{0000-0002-9984-163X}}
\email{chenyanke@pku.edu.cn}
\affiliation{School of Physics and Center of High Energy Physics, Peking University, Beijing 100871, China}

\author{Liang-Zhen Wen\,\orcidlink{0009-0006-8266-5840}}\email{lzwen@pku.edu.cn} 
\affiliation{School of Physics and Center of High Energy Physics, Peking University, Beijing 100871, China}

\author{Lu Meng \orcidlink{0000-0001-9791-7138}}\email{lmeng@seu.edu.cn}
\affiliation{School of Physics, Southeast University, Nanjing 211189, China}

\author{Shi-Lin Zhu\,\orcidlink{0000-0002-4055-6906}}\email{zhusl@pku.edu.cn}
\affiliation{School of Physics and Center of High Energy Physics, Peking University, Beijing 100871, China}

\begin{abstract}
We investigate heavy-meson molecules by treating the pion as a dynamical degree of freedom. We construct an energy-independent Hamiltonian that couples the two- and three-body Fock spaces. The two-body sector contains a pseudoscalar meson and a vector meson $H_1H_2^{*}$, while the three-body sector consists of two pseudoscalar mesons and a pion $H_1H_2 \pi$. The $H^*$ self-energy, the $H_1H_2^{*}$ pion-exchange interaction, and the $H_1H_2\pi$ three-body propagation are consistently generated through the coupling between the two Fock sectors. We use the Gaussian expansion and complex scaling methods to solve the coupled Schr\"odinger equation. Previous approaches usually reduce three-body effects to energy-dependent two-body kernels and solve the equations iteratively. Our approach obtains the poles by solving a single linear generalized eigenvalue problem. The eigenvectors also provide explicit wave functions in both Fock sectors, allowing a qualitative analysis of channel weights and spatial structure. We benchmark the framework in $T_{cc}^{+}$ and $X(3872)$. The benchmark results are consistent with the current experimental pole determinations. We then apply this formalism to the $BB\pi$, $B\bar{B}\pi$, $DB\pi$ and $D\bar{B}\pi$ systems. We predict one isoscalar bound state in each of these systems. The $DB\pi$ and $D\bar{B}\pi$ systems also contain resonances lying below and above their three-body threshold, respectively. We propose complementary searches for these resonances through $DB^*\to DB\gamma$ and $D\bar B\pi$. This coupled Fock-space framework can be readily extended to other few-body systems in which unstable constituents can decay and change the particle number.
\end{abstract}

\maketitle

\emph{Introduction}---%
The $\Xstate$, discovered by the Belle Collaboration in 2003~\cite{Belle:2003nnu}, opened the era of exotic hadron spectroscopy. Its quantum numbers are $J^{PC}=1^{++}$~\cite{ParticleDataGroup:2026mpi}. Recently, BESIII extracted its near-threshold $T$-matrix pole from a coupled-channel line-shape analysis~\cite{BESIII:2023hml}. In 2021, LHCb observed $\Tcc$ in the $D^0D^0\pi^+$ spectrum~\cite{LHCb:2021vvq}. A subsequent line-shape analysis determined its pole below the $D^0D^{*+}$ threshold~\cite{LHCb:2021auc}. These two states have attracted extensive theoretical and experimental interest and revitalized hadron physics. See Refs.~\cite{Chen:2016qju,Esposito:2016noz,Hosaka:2016pey,Lebed:2016hpi,Ali:2017jda,Guo:2017jvc,Olsen:2017bmm,Karliner:2017qhf,Liu:2019zoy,Brambilla:2019esw,Chen:2022asf,Meng:2022ozq,Wu:2022ftm,Mai:2022eur,Liu:2024uxn} for recent reviews.

The proximity of both states to $S$-wave heavy-meson thresholds makes the molecular interpretation natural. The one-pion-exchange (OPE) model offers a simple and physically intuitive framework. Pion exchange can provide the long-range interaction between heavy mesons, while short-range dynamics can be absorbed into cutoff-dependent contact terms. OPE is also a cornerstone of the chiral nuclear forces~\cite{Weinberg:1990rz,Weinberg:1991um}. In heavy meson systems, its form is determined by the heavy-meson chiral Lagrangian~\cite{Wise:1992hn,Casalbuoni:1996pg}. Boson-exchange models involving pion exchange have successfully described the near-threshold $D^{(*)}\bar D^{(*)}$ and $D^{(*)}D^{(*)}$ systems~\cite{Li:2012cs,Li:2012ss,Tornqvist:2004qy,Cheng:2022qcm,Lin:2022wmj,Wang:2024ytk,Chen:2026ohg}.  Notably, boson-exchange calculations predicted an isoscalar doubly charmed molecule before the discovery of $\Tcc$~\cite{Li:2012ss}. The $B^{(*)} B^{(*)}, B^{(*)} \bar{B}^{(*)}, D^{(*)} B^{(*)}$, and $D^{(*)} \bar{B}^{(*)}$ molecules have been investigated with boson-exchange models and contact interactions~\cite{Ding:2009vj,Li:2012ss,Abreu:2015jma,Song:2025ijd}. Recently, $P$-wave di-meson systems have also been investigated within boson-exchange models~\cite{Lin:2024qcq,Cheng:2026cgo,Wang:2024ukc,Wang:2025kpm}.

The pion exchange also connects the two-body interaction to the three-body effects. For the $HD^*$ systems, $D^*\to D\pi$ is allowed with $m_{D^*}-m_D>m_\pi$. The exchanged pion can be on shell, and the open $HD\pi$ channel generates a three-body right-hand cut. The pion exchange and the constituent decay therefore require a consistent treatment. Dynamical pions and three-body effects have been studied in $\Xstate$ and $\Tcc$~\cite{Baru:2011rs,Schmidt:2018vvl,Du:2021zzh,Cheng:2022qcm,Lin:2022wmj,Wang:2023iaz,Hansen:2024ffk,Dawid:2024dgy}. Previous approaches usually reduce three-body effects to energy-dependent two-body kernels and solve the equations iteratively. Effective two-body descriptions also leave the three-body degrees of freedom implicit. A simple and intuitive treatment of these ingredients is our central motivation. An energy-independent Hamiltonian in an enlarged Fock space can describe the same low-energy physics~\cite{Li:2022aru,Yu:2023xxf}.

In this Letter, we construct an energy-independent Hamiltonian for heavy-meson molecules with explicit dynamical pions. It couples the two- and three-body Fock spaces. The constituent self-energy, pion-exchange interaction, and three-body propagation are consistently generated through the coupling between the two Fock sectors. The Gaussian expansion method (GEM) and complex scaling method (CSM) give a single linear generalized eigenvalue problem. Its solutions provide the poles and explicit wave functions in both Fock sectors.  The wave functions also provide qualitative channel weights and spatial structures. Our framework can be extended to other few-body systems in which unstable constituents can decay and change the particle number. We first benchmark the framework in $\Tcc$ and $\Xstate$. We then apply this formalism to make predictions in the bottom sectors.

\emph{Dynamical-pion Hamiltonian}---%
The state vector and the Hamiltonian are defined on the direct sum of the two Fock sectors
\begin{equation}
\hat{\mathcal H}\begin{pmatrix}\Psi_2\\\Psi_3\end{pmatrix}
=E\begin{pmatrix}\Psi_2\\\Psi_3\end{pmatrix},\qquad
\hat{\mathcal H}=\begin{pmatrix}H_{22}&W_{23}\\W_{32}&H_{33}\end{pmatrix}.
\label{eq:prl-hamiltonian}
\end{equation}
The two-body sector contains an $S$-wave pseudoscalar--vector pair. The three-body sector contains two pseudoscalars and a pion with total orbital angular momentum $L=1$.
The three-body block contains only the free propagation of the three constituents,
\begin{equation}
\begin{aligned}
H_{33}&=2m_D+m_\pi+\frac{p_1^2+p_2^2}{2m_D}+T_\pi(p_\pi),
\end{aligned}
\label{eq:h33}
\end{equation}
where $\bm p_1$, $\bm p_2$, and $\bm p_\pi$ are the particle momenta in the total center-of-mass frame. We compare the nonrelativistic (NR) and relativistic (R) pion kinetic energies, 
\begin{equation}
T_\pi^{\NR}(p)=\frac{p^2}{2m_\pi},\quad
T_\pi^{\Rpi}(p)=\sqrt{m_\pi^2+p^2}-m_\pi,
\label{eq:pionkinetic}
\end{equation}
referred to as the NR $\pi$ and R $\pi$ schemes below. The $W_{32}$ converts a vector meson into a pseudoscalar and a pion and $W_{23}=W_{32}^{\dagger}$. The leading heavy-meson chiral interaction gives the internal $P$-wave vertex~\cite{Lin:2022wmj}:
\begin{equation}
W_{32}(\bm k)=\lambda_{0,\Lambda}\tau^a
\frac{\bm\epsilon\!\cdot\!\bm k~e^{-k^2/\Lambda^2}}{\sqrt{2\omega_k}}
,\quad \omega_k=\sqrt{m_\pi^2+k^2},
\label{eq:prl-vertex}
\end{equation}
after factoring out the $\delta$-function for momentum conservation. The momentum $\bm k$ is internal to the daughter heavy-meson--pion pair, $\bm\epsilon$ is the vector polarization, $\tau^a$ fixes the isospin coefficients, $\Lambda$ is the cutoff, and $\lambda_{0,\Lambda}=g_{0,\Lambda}/(2f_{\pi})$ is the reduced bare axial coupling. The transition sums all allowed daughter pairings.

\begin{figure}[tb]
\centering
\includegraphics[width=0.9\columnwidth]{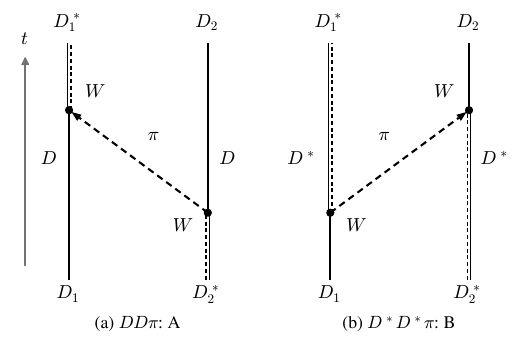}
\caption{The two time orderings of pion exchange in $DD^*$, with time increasing upward. Solid--dashed double lines denote bare $D^*$, solid lines denote $D$, and dashed lines denote pions. The vertices are labeled $W$. (a) The $DD\pi$ intermediate state is included explicitly through $W_{32}$ and $H_{33}$. (b) The $D^*D^*\pi$ intermediate state gives the static interaction in $H_{22}$.}
\label{fig:timeorderings}
\end{figure}

The explicit dynamical pion connects the exchange interaction to the opening of the three-body channel. Pion emission and absorption on the same heavy-meson line dress the vector constituent, and pion absorption on the other line generates the pion exchange interaction. Both processes use the same three-body propagation and transition vertex $W$. Figure~\ref{fig:timeorderings} illustrates two time orderings of pion exchange in $DD^*$. The low-energy pion time ordering A with the $DD\pi$ intermediate state is included in Eq.~\eqref{eq:prl-hamiltonian}. The complementary higher-energy time ordering is represented by a static interaction $V_B^S$ in $H_{22}$. This assignment includes each time ordering once. Thus the two-body block reads
\begin{equation}
H_{22}=m_D+m_{D^*,0,\Lambda}
+\frac{p^2}{2m_D}+\frac{p^2}{2m_{D^*,0,\Lambda}}+V_B^S+V_C,
\label{eq:h22}
\end{equation}
where $m_{D^*,0,\Lambda}$ is the bare vector mass and $V_C$ is the short-range contact interaction. We calculate the $D\pi$ self-energy and fix the bare vector mass and coupling by matching the dressed $D^*$ pole to its physical mass and strong width. For $D^{*+}$, the strong width input is $82.1~\mathrm{keV}$ after removal of its radiative branch~\cite{BaBar:2013zgp,ParticleDataGroup:2026mpi}. The same coupling predicts the neutral $D^{*0}$ strong width. Once the parameters are calibrated, all Hamiltonian blocks are energy-independent. The $D^*$ width arises from the transition vertex $W$ itself. The detailed interactions and the charge coefficients are given in the Supplemental Material~\cite{Suppl}.

We solve the Schr\"odinger equation with the Gaussian expansion method (GEM)~\cite{Hiyama:2003cu} and use the complex scaling method (CSM)~\cite{Aguilar:1971ve,Balslev:1971vb,Moiseyev:1998gjp,Aoyama:2006hrz,Chen:2023eri} to identify genuine resonances. The CSM rotates all intrinsic coordinates and momenta by $r\to re^{i\theta}$ and $p\to pe^{-i\theta}$. Bound states and resonances remain stationary as $\theta$ changes, while scattering states rotate from their thresholds. The combination of the GEM and the CSM has proven to be a powerful tool for investigating few-body resonances~\cite{Aoyama:2006hrz,Carbonell:2013ywa,Hiyama:2016nwn,Dote:2017wkk,Happ:2023kcc,Chen:2023eri,Chen:2023syh,Meng:2024yhu,Ma:2024vsi,Wu:2024euj,Wu:2024zbx,Zheng:2025uzy,Wen:2026rsy}. The basis functions, complex scaling conventions, and pole identification scheme are provided in detail in the Supplemental Material~\cite{Suppl}.

\begin{table}[t]
\centering
\caption{Benchmark poles and fitted contacts at $\Lambda=750~\mathrm{MeV}$ with NR $\pi$. Daggers mark input masses. The offsets $\delta M$ use thresholds of $3875.110$ and $3871.700~\mathrm{MeV}$ for $\Tcc$ and $\Xstate$, respectively. Experimental inputs are from Refs.~\cite{LHCb:2021auc,BESIII:2023hml}.}
\label{tab:poles750nr}
\begin{tabular*}{\columnwidth}{@{}@{\extracolsep{\fill}}lcc@{}}
\hline\hline
Quantity & Theoretical & Experimental\\
\hline
\multicolumn{3}{c}{$\Tcc$}\\
$C_0$ ($\mathrm{GeV}^{-2}$) & $-28.3886$ & ---\\
$M$ (MeV) & $3874.750^{\dagger}$ & $3874.750\pm0.040{}^{+0.004}_{-0}$\\
$\delta M$ (keV) & $-360$ & $-360\pm40{}^{+4}_{-0}$\\
$\operatorname{Im}[\epole]$ (keV) & $-26$ & $-24\pm1{}^{+7}_{-0}$\\
$\Gamma$ (keV) & $53$ & $48\pm2{}^{+0}_{-14}$\\
\hline
\multicolumn{3}{c}{$\Xstate$}\\
$C_0^X$ ($\mathrm{GeV}^{-2}$) & $-31.5471$ & ---\\
$M$ (MeV) & $3871.600^{\dagger}$ & $3871.70\pm0.15{}^{+0.07}_{-0.08}$\\
$\delta M$ (keV) & $-100$ & $0\pm150{}^{+70}_{-80}$\\
$\operatorname{Im}[\epole]$ (keV) & $-19$ & $-190\pm80{}^{+140}_{-190}$\\
$\Gamma$ (keV) & $38$ & $380\pm160{}^{+380}_{-280}$\\
\hline
\hline
\end{tabular*}
\end{table}

\begin{figure*}[htbp]
\centering
\includegraphics[width=\textwidth]{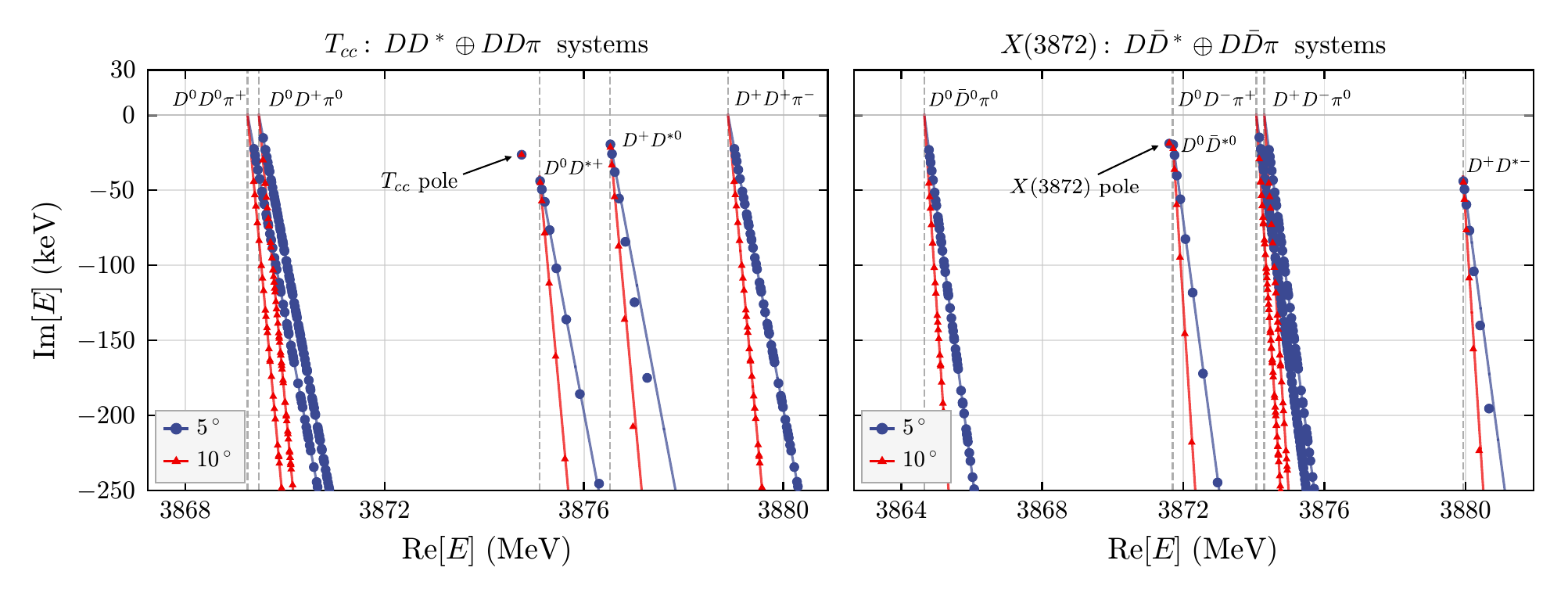}
\caption{CSM spectra of the coupled $DD^*\!\oplus DD\pi$ system for $\Tcc$ and $D\bar D^*\!\oplus D\bar D\pi$ system for $\Xstate$ with $\Lambda=750~\mathrm{MeV}$ and NR $\pi$. Blue filled circles and red filled triangles are eigenvalues for $\theta=5^\circ$ and $10^\circ$. Solid lines in the same colors indicate the corresponding threshold rays. Three-body rays start from real thresholds, and the two-body rays start from the dressed $D^*$ pole plus the spectator mass.}
\label{fig:csm}
\end{figure*}

\begin{figure*}[htbp]
\centering
\includegraphics[width=0.8\textwidth]{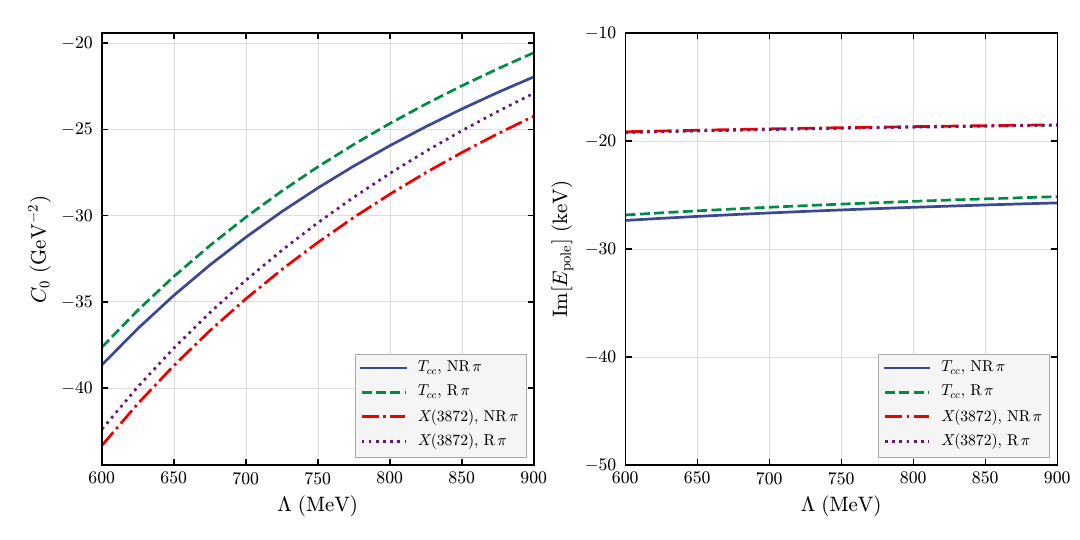}
\caption{Cutoff dependence of the fitted isoscalar contact and the predicted pole imaginary part. The pole real parts are fixed at $3874.750$ and $3871.600~\mathrm{MeV}$ for $\Tcc$ and $\Xstate$. Blue solid, green dashed, red dash-dotted, and purple dotted curves denote $\Tcc$ with NR $\pi$, $\Tcc$ with R $\pi$, $\Xstate$ with NR $\pi$, and $\Xstate$ with R $\pi$, respectively.}
\label{fig:lambda}
\end{figure*}

\emph{Charm benchmarks}---%
For $\Tcc$ and $\Xstate$, we include physical charge masses and fit isoscalar contacts $C_0$ and $C_0^X$ to the pole masses. Their widths are obtained as predictions and compared with experiment. Figure~\ref{fig:csm} shows the CSM spectra of the coupled $DD^*\oplus DD\pi$ system for $\Tcc$ and the $D\bar D^*\oplus D\bar D\pi$ system for $\Xstate$. The rays from the real three-body thresholds and from the complex $D^*$ pole positions are resolved simultaneously. With $\Lambda=750~\mathrm{MeV}$ and NR $\pi$ kinetic energy, we obtain
\begin{equation}
\begin{aligned}
E_{\mathrm{pole}}^{T_{cc}}&=3874.750-i\,0.026~\mathrm{MeV},\\
E_{\mathrm{pole}}^{X}&=3871.600-i\,0.019~\mathrm{MeV}.
\end{aligned}
\label{eq:prl-charm}
\end{equation}
For $\Xstate$, we search for pole masses at both $3871.60$ and $3871.80~\mathrm{MeV}$, which are $100~\mathrm{keV}$ below and above the $D^0\bar{D}^{*0}$ threshold, respectively. The $3871.60~\mathrm{MeV}$ target produces a pole stable under the scaling angles. The $3871.80~\mathrm{MeV}$ search does not yield an acceptable solution. The current framework therefore favors a pole below the $D^0\bar D^{*0}$ threshold. Table~\ref{tab:poles750nr} compares these poles with experiment and lists the fitted contacts. The predicted widths are consistent with the experimental values. The calculated poles also reproduce the experimental Riemann-sheet assignments~\cite{LHCb:2021auc,BESIII:2023hml}. 

The benchmark results are essentially independent of the cutoff and the treatment of the pion kinetic energy. Figure~\ref{fig:lambda} shows a cutoff scan from $600$ to $900~\mathrm{MeV}$. At each cutoff, the $D^*$ bare parameters are recalibrated to their physical inputs and the real isoscalar contact is refitted to the pole masses. The contact runs smoothly with $\Lambda$. The pole imaginary part changes little. This framework also predicts a narrower $\Xstate$ than $\Tcc$ throughout the cutoff scan and for both pion kinetic schemes. 

This hierarchy has a direct origin. At the adopted masses, $\Tcc$ has two open $DD\pi$ charge channels, whereas $\Xstate$ has only the $D^0\bar D^0\pi^0$ channel. The available channels and their charge-weighted phase space directly explain this width hierarchy. A similar width hierarchy was also obtained in Ref.~\cite{Lin:2022wmj}. The detailed parameter scans, wave function compositions and spatial structures are presented in the Supplemental Material~\cite{Suppl}.

\begin{figure*}[htbp]
\centering
\includegraphics[width=\textwidth]{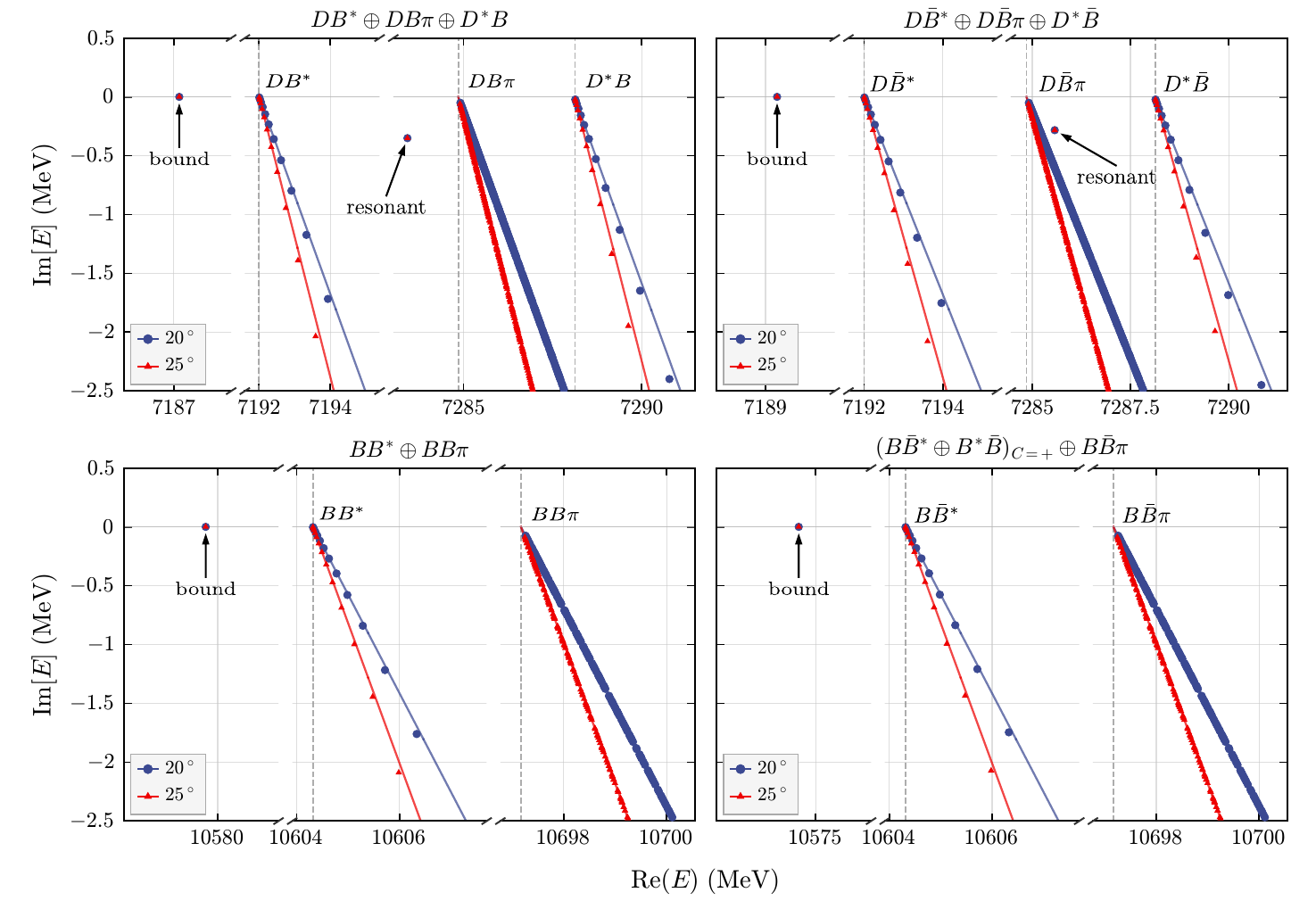}
\caption{CSM spectra of the $cb$ and $c\bar{b}$ systems with $\Lambda=750~\mathrm{MeV}$ and NR $\pi$. Blue circles and red triangles are eigenvalues at $\theta=20^\circ$ and $25^\circ$. Solid lines indicate threshold rays. The resonances lie on opposite sides of the three-body threshold.}
\label{fig:prl-spectra}
\end{figure*}
\begin{figure*}[htbp]
\centering
\includegraphics[width=\textwidth]{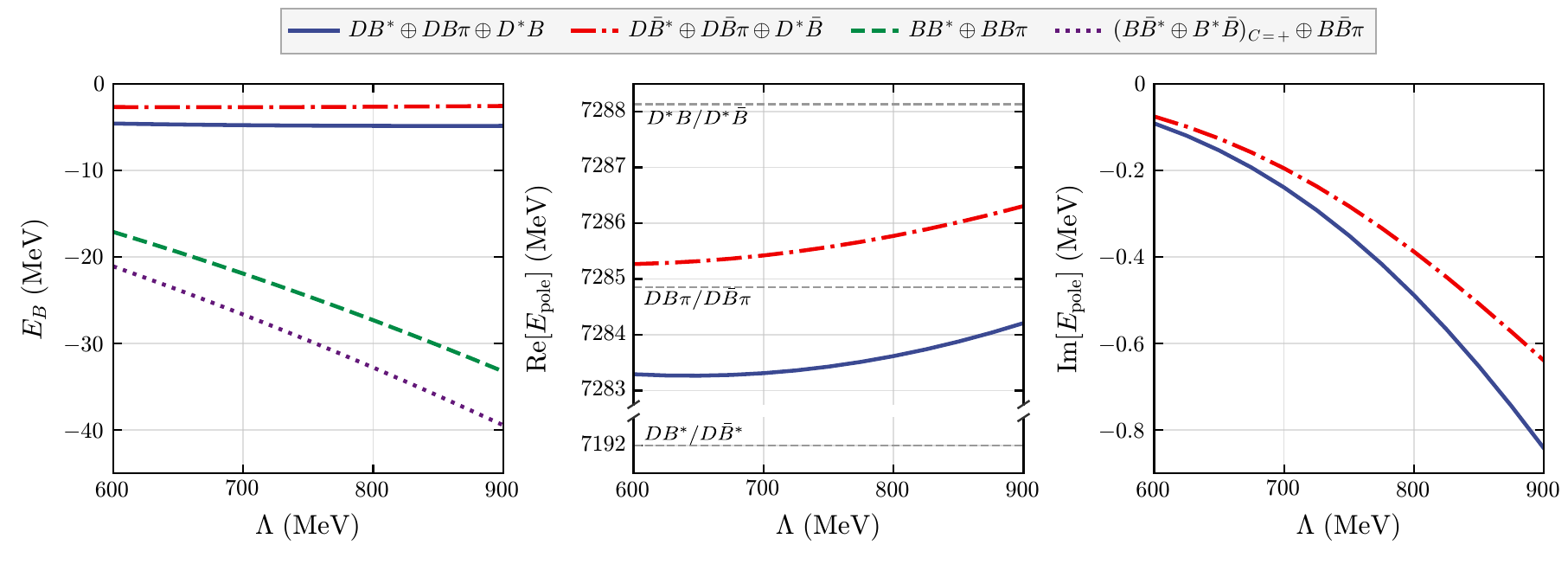}
\caption{Cutoff dependence of the four isoscalar $cb$, $c\bar{b}$, $bb$ and $b\bar{b}$ systems. Left: the binding energy $E_B=M-E_{\mathrm{th}}^{\mathrm{low}}$ of the four lowest states. Middle: the real parts of the $cb$ and $c\bar{b}$ resonance poles. Right: the imaginary parts of the $cb$ and $c\bar{b}$ resonance poles.}
\label{fig:prl-cutoff}
\end{figure*}

\emph{Predictions in bottom systems}---%
We extend the same Hamiltonian to four isoscalar systems containing bottom mesons. We label the systems by their heavy-quark contents. The $c\bar b$, $cb$, and $bb$ systems contain the coupled channels $DB^*\oplus DB\pi\oplus D^*B$, $D\bar B^*\oplus D\bar B\pi\oplus D^*\bar B$, and $BB^*\oplus BB\pi$, respectively, and have $I(J^P)=0(1^+)$. The $b\bar b$ system contains $(B\bar B^*\oplus B^*\bar B)_{C=+}\oplus B\bar B\pi$ and has $I(J^{PC})=0(1^{++})$.

According to the heavy-quark symmetry, the physical axial coupling and contact low-energy constants should be flavor independent at leading order~\cite{Neubert:1993mb}. We take the same physical axial coupling $\lambda_B=\lambda_D$ and
\begin{equation}
\begin{aligned}
C_{BB}&=C_0, & C_{B\bar B}&=C_0^X,\\
\mathbf C_{D\bar B}&=C_0\mathbf1_2, &
\mathbf C_{DB}&=C_0^X\mathbf1_2.
\end{aligned}
\label{eq:bottomcontacts}
\end{equation}
For the $(DB^*,D^*B)$ and $(D\bar B^*,D^*\bar B)$ systems, we neglect the off-diagonal contact for the two channels separated by $96.13~\mathrm{MeV}$. Their off-diagonal interaction is generated by pion exchange. Both the explicit low-energy time ordering and the static complementary term contribute to this interaction.

The thresholds of the $cb(\bar{b})$ systems are~\cite{ParticleDataGroup:2026mpi}:
\begin{equation}
\begin{aligned}
m_D+m_{B^*}&=7192.000~\mathrm{MeV},\\
m_D+m_B+m_\pi&=7284.854~\mathrm{MeV},\\
m_{D^*}+m_B&=7288.130~\mathrm{MeV}.
\end{aligned}
\label{eq:prl-thresholds}
\end{equation}
The first channel contains a $B^*$ stable against strong decay, while the highest nominal threshold contains the unstable $D^*$. The intervening three-body threshold distinguishes the two regimes.

Figure~\ref{fig:prl-spectra} shows the complex-scaled spectra. A bound state appears below the lowest two-body threshold in each system. The $cb$ and $c\bar b$ systems have binding energies of several MeV, while the $bb$ and $b\bar{b}$ systems bind by several tens of MeV. Shallow $cb$ bound states are also supported by lattice QCD and a subsequent EFT analysis~\cite{Alexandrou:2023cqg,Abolnikov:2026sik}.

The $cb$ and $c\bar{b}$ systems also contain resonances near the higher two-body thresholds. At $\Lambda=750~\mathrm{MeV}$, their poles are
\begin{equation}
\begin{aligned}
E_{\mathrm{pole}}^{c\bar b}&=7283.429-i\,0.351~\mathrm{MeV},\\
E_{\mathrm{pole}}^{cb}&=7285.573-i\,0.283~\mathrm{MeV}.
\end{aligned}
\label{eq:prl-mixedpoles}
\end{equation}
The $c\bar b$ pole lies $1.425~\mathrm{MeV}$ below the three-body threshold, and the $cb$ pole lies $0.718~\mathrm{MeV}$ above it. Both resonances can decay to their lower two-body channels. The $cb$ resonance can also decay to the $D\bar{B}\pi$ channel. Pion emission and absorption couple the two two-body channels. In particular, with zero off-diagonal contact, the decay of the quasi-bound resonance into the lower channel proceeds solely through the mediated pion. The resonance positions and widths are determined together by this dynamics. The explicit dynamical pion makes the relation between channel conversion and the three-body threshold visible in the same spectrum.

The detailed parameter scans, wave function compositions and spatial structures are presented in the Supplemental Material~\cite{Suppl}. Figure~\ref{fig:prl-cutoff} shows the cutoff dependence over $600$--$900~\mathrm{MeV}$. The $c\bar b$ and $cb$ states remain shallowly bound, with nearly constant binding energies. The $bb$ and $b\bar b$ states are more deeply bound and more sensitive to the cutoff. This behavior is readily understood from the reduced masses. For the $c\bar b$ and $cb$ systems, the reduced mass $\mu_{DB^*}=m_Dm_{B^*}/(m_D+m_{B^*})$ is controlled mainly by the lighter $D$ meson. The larger reduced masses of the $bb$ and $b\bar b$ systems suppress the kinetic energy, making these states more sensitive to the short-range interaction.

The two resonance masses each vary by about $1~\mathrm{MeV}$ across the cutoff range. The $c\bar b$ pole remains below the three-body threshold and the $cb$ pole remains above the three-body threshold throughout. Their imaginary parts are more cutoff-sensitive. The changing open-channel phase space contributes to the width variation. 

The lowest $bb$ and $cb$ states are stable against strong decay. The weak decays and radiative decays to $BB\gamma$ and $D\bar B\gamma$ are allowed. The $c\bar b$ and $b\bar b$ states can also decay through heavy-quark rearrangement, for example into $B_c\omega$ or $B_c^*\eta$ for $c\bar b$, and $\Upsilon(1S)\omega$ for $b\bar b$. The couplings to the rearrangement channels and their decay widths are beyond the present calculation.

We also propose complementary searches for the two $cb(\bar{b})$ resonances. For the $cb$ resonance, a search can be conducted using the $D\bar{B}\pi$ final state. For the $c\bar{b}$ resonance, the $DB^* \to DB\gamma$ final state can be utilized. The recent LHCb observation of $B_{s0}^*(5700)^0\to B_s^0\pi^0$ uses $\pi^0\to\gamma\gamma$ reconstructed as two separate clusters in the electromagnetic calorimeter~\cite{LHCb:2026knl}. This demonstration of photon reconstruction in bottom-hadron spectroscopy supports searches for the predicted threshold pattern in pion and photon final states.

\emph{Summary}---%
An explicit dynamical pion connects molecular exchange, constituent dressing, and open three-body propagation in a energy-independent Hamiltonian. GEM and CSM provide a common linear spectral calculation for stable and unstable vector constituents. We benchmark the framework in $\Tcc$ and $\Xstate$ systems. The results are in good agreement with experiment and depend weakly on the cutoff. We predict one isoscalar bound state in each of the $cb$, $c\bar{b}$, $bb$ and $b\bar{b}$ sectors. The $cb(\bar{b})$ systems bind by several MeV with weak cutoff dependence. The $bb(\bar{b})$ binding energies are several tens of MeV. We also predict two $cb(\bar{b})$ resonances lying on opposite sides of the three-body threshold throughout $\Lambda=600$--$900~\mathrm{MeV}$. We propose searches through $DB^*\to DB\gamma$ for the $c \bar{b}$ resonance and $D \bar{B} \pi$ for the $c b$ resonance. The coupled Fock-space construction supplies a unified treatment of the heavy vector meson self-energy, the long-range pion exchange and three-body propagation. Our framework is not restricted to two- and three-body Fock sectors or to heavy-meson molecules. It can be readily extended to other few-body systems with unstable constituents and variable particle number.

\emph{Acknowledgments}---%
We thank Zi-Yang Lin for helpful discussions. This project was supported by the National Natural Science Foundation of China (Grants No. 12475137 and No. 12605136) and Start-up Funds of Southeast University (Grant No. 4007022506). The computational resources were supported by the High-performance Computing Platform of Peking University.

\emph{Data availability}---The numerical values and figures supporting the conclusions are presented in this Letter and its Supplemental Material. The underlying spectra, parameter scans, and channel wave functions are available from the authors upon request.

\bibliography{references}
\clearpage
\onecolumngrid
\begin{center}
\large\bfseries Supplemental Material
\end{center}
\setcounter{section}{0}
\setcounter{subsection}{0}
\renewcommand{\theHsection}{S.\arabic{section}}
\renewcommand{\theHsubsection}{S.\arabic{section}.\arabic{subsection}}
\renewcommand{\theHequation}{S.\arabic{equation}}
\renewcommand{\theHfigure}{S.\arabic{figure}}
\renewcommand{\theHtable}{S.\arabic{table}}
\setcounter{secnumdepth}{3}
\renewcommand{\thesection}{S\arabic{section}}
\renewcommand{\thesubsection}{\thesection.\Alph{subsection}}
\makeatletter
\renewcommand{\p@subsection}{}
\renewcommand{\p@subsubsection}{}
\makeatother
\renewcommand{\theequation}{S\arabic{equation}}
\renewcommand{\thefigure}{S\arabic{figure}}
\renewcommand{\thetable}{S\Roman{table}}
\setcounter{equation}{0}
\setcounter{figure}{0}
\setcounter{table}{0}
\begingroup\setlength{\fboxsep}{6pt}
\noindent\begin{minipage}{\dimexpr\linewidth-2\fboxsep-2\fboxrule\relax}
This Supplemental Material collects the technical details not elaborated upon in the Letter. 
Equations, tables, and figures below carry an S prefix. References use the numbering of the Letter.
\end{minipage}\endgroup\par\medskip

\section{Hamiltonian}\label{sec:hamiltonian}

For compactness, the formulas in this section are written in the isospin limit. In the actual numerical calculations for $\Tcc$ and $\Xstate$, we use physical charge masses. Table~\ref{tab:masses} lists the physical charge masses used in the calculation~\cite{ParticleDataGroup:2026mpi}. Table~\ref{tab:thresholds} collects the charge channel thresholds. For the isospin-limit formulas, we use
\begin{equation}
m_D=1867.25~\mathrm{MeV},\qquad m_{D^*}=2008.565~\mathrm{MeV},\qquad m_\pi=138.039~\mathrm{MeV}.
\label{eq:isomasses}
\end{equation}
We consider $I(J^P)=0(1^+)$ for $\Tcc$ and $I(J^{PC})=0(1^{++})$ for $\Xstate$. For the $C=+$ two-body states, we use the field phases $C(D)=\bar D$ and $C(D^*)=-\bar D^*$, giving
\begin{equation}
|n\rangle=\frac{|D^0\bar D^{*0}\rangle-|D^{*0}\bar D^0\rangle}{\sqrt2},\qquad
|c\rangle=\frac{|D^+D^{*-}\rangle-|D^{*+}D^-\rangle}{\sqrt2}.
\label{eq:cchannels}
\end{equation}
Both states have $C=+$. The extension to the bottom sector is given in Sec.~\ref{sec:bottom}.

\begin{table}[htbp]
\centering
\caption{Physical charge masses used in the calculation, in MeV~\cite{ParticleDataGroup:2026mpi}. Particle and antiparticle masses are equal.}
\label{tab:masses}
\begin{tabular*}{\columnwidth}{@{\extracolsep{\fill}}lrrrrrr@{}}
\hline\hline
Particle & $D^0$ & $D^+$ & $D^{*0}$ & $D^{*+}$ & $\pi^\pm$ & $\pi^0$\\
\hline
Mass & $1864.84$ & $1869.66$ & $2006.86$ & $2010.27$ & $139.570$ & $134.977$\\
\hline\hline
\end{tabular*}
\end{table}

\begin{table}[htbp]
\centering
\caption{Charge channels and their thresholds. For the unstable $D^*$, only its mass is used to calculate the listed thresholds.}
\label{tab:thresholds}
\begin{tabular*}{\columnwidth}{@{\extracolsep{\fill}}llc@{}}
\hline\hline
System & Channel & Threshold (MeV)\\
\hline
$\Tcc$ & $D^0D^0\pi^+$ & 3869.250\\
 & $D^0D^+\pi^0$ & 3869.477\\
 & $D^0D^{*+}$ & 3875.110\\
 & $D^+D^{*0}$ & 3876.520\\
 & $D^+D^+\pi^-$ & 3878.890\\
\hline
$\Xstate$ & $D^0\bar D^0\pi^0$ & 3864.657\\
 & $D^0\bar D^{*0},\ D^{*0}\bar D^0$ & 3871.700\\
 & $D^0D^-\pi^+,\ D^+\bar D^0\pi^-$ & 3874.070\\
 & $D^+D^-\pi^0$ & 3874.297\\
 & $D^+D^{*-},\ D^{*+}D^-$ & 3879.930\\
\hline\hline
\end{tabular*}
\end{table}

\subsection{Chiral interactions}\label{sec:chiral}
Following Ref.~\cite{Lin:2022wmj}, the leading heavy-meson chiral Lagrangian reads
\begin{align}
\mathcal L_{H\pi}^{(1)}={}&-\langle(iv\!\cdot\!\partial H)\bar H\rangle
+\langle H v\!\cdot\!\Gamma\bar H\rangle+g\langle H\slashed u\gamma_5\bar H\rangle
-\frac{\Delta}{8}\langle H\sigma^{\mu\nu}\bar H\sigma_{\mu\nu}\rangle.\\
\mathcal L_{\tilde H\pi}^{(1)}={}&-\langle(iv\!\cdot\!\partial \bar{\tilde{H}})\tilde H\rangle
+\langle \bar{\tilde{H}} v\!\cdot\!\Gamma\tilde H\rangle+g\langle \bar{\tilde{H}}\slashed u\gamma_5\tilde H\rangle
-\frac{\Delta}{8}\langle \bar{\tilde{H}}\sigma^{\mu\nu}\tilde H\sigma_{\mu\nu}\rangle.
\label{eq:lagrangian}
\end{align}
Here $\langle\cdots\rangle$ denotes the trace, $v=(1,\bm0)$ is the heavy-meson velocity, $g$ is the axial coupling, and $\Delta=m_{D^*}-m_D$. The heavy spin multiplet and the axial field are
\begin{align}
H&=\frac{1+\slashed v}{2}(D^*_{\mu}\gamma^\mu+iD\gamma_5),
\qquad \bar H=\gamma^0H^\dagger\gamma^0,\nonumber\\
\tilde{H}&=(\tilde{D}_\mu^* \gamma^\mu+i \tilde{D} \gamma_5) \frac{1-\slashed v}{2},\qquad \bar{\tilde{H}}=\gamma^0 \tilde H^{\dagger} \gamma^0, \nonumber \\
u_\mu&=-\frac{\tau^a\partial_\mu\pi^a}{2f_\pi}+\cdots,
\qquad \xi=\exp\!\left(\frac{i\tau^a\pi^a}{2f_\pi}\right),
\label{eq:heavyfield}
\end{align}
where $D^{(*)}=(D^{(*)0},D^{(*)+})$ and $\tilde D^{(*)}=(\bar{D}^{(*)0},\bar{D}^{(*)+})^T$ are the heavy meson isospin doublets, $\tau^a$ are the Pauli matrices, and $f_\pi$ is the pion decay constant. The chiral connection $\Gamma_\mu=(i/2)[\xi^\dagger,\partial_\mu\xi]$ starts with two pion fields.

The leading four-heavy-meson contact Lagrangian reads
\begin{align}
\mathcal L_{HH}^{(0)}={}&D_a\langle H\gamma_\mu\bar H\rangle
\langle H\gamma^\mu\bar H\rangle+D_b\langle H\gamma_\mu\gamma_5\bar H\rangle
\langle H\gamma^\mu\gamma_5\bar H\rangle\nonumber\\
&+E_a\langle H\gamma_\mu\tau^a\bar H\rangle
\langle H\gamma^\mu\tau^a\bar H\rangle+E_b\langle H\gamma_\mu\gamma_5\tau^a\bar H\rangle
\langle H\gamma^\mu\gamma_5\tau^a\bar H\rangle,\\
\mathcal L_{H\tilde H}^{(0)}={}&2\tilde D_a
\langle\bar{\tilde H}\gamma_\mu\tilde H\rangle
\langle H\gamma^\mu\bar H\rangle+2\tilde D_b\langle\bar{\tilde H}\gamma_\mu\gamma_5\tilde H\rangle
\langle H\gamma^\mu\gamma_5\bar H\rangle\nonumber\\
&+2\tilde E_a\langle\bar{\tilde H}\gamma_\mu\tau^a\tilde H\rangle
\langle H\gamma^\mu\tau^a\bar H\rangle+2\tilde E_b\langle\bar{\tilde H}\gamma_\mu\gamma_5\tau^a\tilde H\rangle
\langle H\gamma^\mu\gamma_5\tau^a\bar H\rangle.
\label{eq:contactlag}
\end{align}
Here $D_{a,b},E_{a,b}$ and their tilded counterparts are independent low-energy constants. Their $S$-wave projections onto the $1^+$ $DD^*$ and the $1^{++}$ $D\bar D^*$ channels give
\begin{align}
C_0&=-2(D_a+D_b)+6(E_a+E_b),\nonumber\\
C_1&=-2(D_a-D_b)-2(E_a-E_b),\nonumber\\
C_0^X&=2(\tilde D_a+\tilde D_b)+6(\tilde E_a+\tilde E_b),\nonumber\\
C_1^X&=2(\tilde D_a+\tilde D_b)-2(\tilde E_a+\tilde E_b).
\label{eq:contactLECs}
\end{align}
The subscripts $0$ and $1$ denote isoscalar and isovector channels, respectively.

\subsection{Transition vertex and constituent pole}\label{sec:vertex-calibration}

For a $D_jD_i^*$ system decaying into the daughter pair $D_i\pi$ with spectator $D_j$, we introduce the Jacobi coordinates
\begin{equation}
\bm x=\bm r_\pi-\bm r_i,\qquad
\bm y=\bm r_j-\frac{m_D\bm r_i+m_\pi\bm r_\pi}{m_D+m_\pi},
\label{eq:vertexjacobi}
\end{equation}
with conjugate momenta $\bm k$ and $\bm p$. The spectator carries $\bm p$ and the daughter pair carries $-\bm p$. In the initial two-body state, $\bm q$ is conjugate to the separation $\bm R$ between the spectator $D_j$ and the $D_i^*$. In the total center-of-mass frame,
\begin{equation}
\begin{aligned}
\bm p_\pi&=\bm k-\alpha_\pi\bm p,\qquad
\bm p_i=-\bm k-\alpha_D\bm p,\\
\alpha_D&=\frac{m_D}{m_D+m_\pi},\qquad\alpha_\pi=1-\alpha_D.
\end{aligned}
\label{eq:movingvertex}
\end{equation}
The full form of the transition vertex quoted in the Letter then reads:
\begin{align}
\langle\bm k,\bm p|W_{32}^{(i)}|\bm q,\bm\epsilon\rangle
=&(2\pi)^3\delta^{(3)}(\bm p-\bm q)\lambda_{0,\Lambda}\tau^a
\frac{\bm\epsilon\cdot\bm k}{\sqrt{2\omega_k}}f_\Lambda(k),
\label{eq:vertex}
\\[-2mm]
\omega_k=&\sqrt{m_\pi^2+k^2},\nonumber\\
f_\Lambda(k)=&e^{-k^2/\Lambda^2}.
\label{eq:regulator}
\end{align}
Here $\bm\epsilon$ is the $D^*$ polarization, $\Lambda$ is the regulator cutoff, and $\lambda_{0,\Lambda}$ is the bare coupling, related to the axial coupling of \eq{eq:lagrangian} by $g_{0,\Lambda}=2f_\pi\lambda_{0,\Lambda}$. The factor $\bm\epsilon\cdot\bm k$ selects the internal $P$ wave of the daughter pair, and the isospin factor couples $D_i\pi$ to $I(J^P)=1/2(1^-)$. The complete transition sums the allowed daughter pairings with their charge coefficients and the Bose or charge-conjugation projection collected in Sec.~\ref{app:charge}. Both pion kinetic schemes retain the same vertex factor $1/\sqrt{2\omega_k}$. This transition has the coordinate-space form
\begin{align}
\langle\bm x,\bm y|W_{32}^{(i)}|\bm R,\bm\epsilon\rangle
={}&\delta^{(3)}(\bm y-\bm R)\times\big[-i\lambda_{0,\Lambda}\tau^a
\bm\epsilon\cdot\bm\nabla_x F_\Lambda(x)\big],
\label{eq:coordinatevertex}
\\
F_\Lambda(x)={}&\int\frac{\dd^3k}{(2\pi)^3}
e^{i\bm k\cdot\bm x}\frac{f_\Lambda(k)}{\sqrt{2\omega_k}}.
\label{eq:coordinateformfactor}
\end{align}
The delta function embeds the initial coordinate $\bm R$ as the spectator coordinate $\bm y$. The derivative acts on the radial function $F_\Lambda(x)$, with $x=|\bm x|$.

We fix the parameters of this vertex using the physical $D^*$ mass and its strong decay width. We use the measured total $D^{*+}$ width $83.4\pm1.8~\mathrm{keV}$ and radiative branching fraction $\mathcal B_\gamma=0.016\pm0.004$~\cite{BaBar:2013zgp,ParticleDataGroup:2026mpi} to obtain the central $D^{*+}$ strong decay width input $\Gamma_{D\pi}^{+}=82.1~\mathrm{keV}$. This strong decay width determines the physical reduced coupling $\lambda_a=g_a/(2f_\pi)$, where $g_a$ is the physical axial coupling. For a charge channel $c$, the leading $P$-wave partial width reads
\begin{equation}
\Gamma_c=C_c\lambda_a^2\frac{m_{D,c}}{m_{D^*}}\frac{k_c^3}{6\pi},
\label{eq:physicalpartialwidth}
\end{equation}
where $m_{D,c}$ is the daughter mass, $m_{D^*}$ the physical parent mass, and $k_c$ the two-body decay momentum. The coefficient $C_c$ is the squared charge matrix element of $\tau^a$ in \eq{eq:vertex}, equal to $2$ and $1$ for $D^{*+}\to D^0\pi^+$ and $D^+\pi^0$. The factor $k_c^3$ combines the squared $P$-wave vertex with the two-body phase space. Using the physical masses and the strong width, we obtain
\begin{equation}
\lambda_a=3.07161~\mathrm{GeV}^{-1},\qquad
\frac{g_a}{f_\pi}=6.14321~\mathrm{GeV}^{-1}.
\label{eq:physicalcoupling}
\end{equation}
We treat $\lambda_a$ as the physical reference and determine the regulated bare coupling $\lambda_{0,\Lambda}$ by pole matching.

\begin{figure}[htbp]
\centering
\includegraphics[width=0.55\columnwidth]{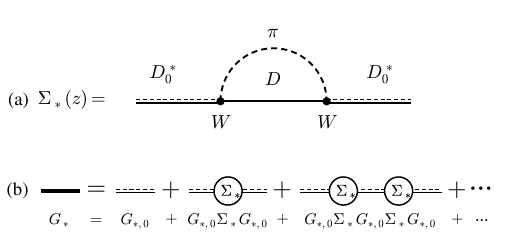}
\caption{$D^*$ self-energy and dressing. (a) Amputated $D\pi$ self-energy loop, with $W$ labeling the regulated bare transition vertex and its Hermitian conjugate. (b) Dyson series for the physical (dressed) propagator $G_*$. Closely spaced solid--dashed double lines denote bare $D^*$ propagators $G_{*,0}$, thick solid lines denote the physical $D^*$, thin solid lines denote $D$, and dashed lines denote pions. Each $\Sigma_*$ insertion corresponds to the loop in (a), summed over charge channels.}
\label{fig:dstardressing}
\end{figure}

Iterating the vertex in \eq{eq:vertex} through the free $D\pi$ subsystem generates the $D^*$ self-energy and dresses the bare propagator, as shown in Fig.~\ref{fig:dstardressing}. In the $D^*$ rest frame, the contribution of charge channel $c$ reads
\begin{equation}
\Sigma_c(z)=\frac{C_c\lambda_{0,\Lambda}^2}{12\pi^2}
\int_0^\infty\dd k\,
\frac{k^4e^{-2k^2/\Lambda^2}}
{\omega_c(k)[z-m_{D,c}-m_{\pi,c}-\varepsilon_c(k)]},
\label{eq:starselfenergy}
\end{equation}
where $z$ is the $D\pi$ subsystem energy. $\varepsilon_c(k)=k^2/(2m_{D,c})+T_{\pi,c}(k)$ follows the kinetic prescription in \eq{eq:pionkinetic} of the Letter. $\omega_c(k)=\sqrt{m_{\pi,c}^2+k^2}$ comes from the common vertex normalization. The factor $k^4$ combines the two $P$-wave vertices with the radial measure, and both vertices are regulated by $f_\Lambda$. The Dyson series gives $G_*(z)=[z-m_{D^*,0,\Lambda}-\Sigma_*(z)]^{-1}$, where $\Sigma_*(z)=\sum_c\Sigma_c(z)$. For each cutoff and pion kinetic scheme, we extract the bare mass $m_{D^*,0,\Lambda}^{+}$ and coupling $\lambda_{0,\Lambda}$ by
\begin{equation}
z_{*,+}=m_{D^*,0,\Lambda}^{+}+\sum_c\Sigma_c(z_{*,+}),\qquad
z_{*,+}=m_{D^{*+}}-\frac{i}{2}\Gamma_{D\pi}^{+}.
\label{eq:starcalibration}
\end{equation}

The neutral $D^{*0}$ bare mass is extracted from the physical $m_{D^{*0}}$ with the same scheme, and its strong decay width $\Gamma_{D\pi}^{0}$ is a predicted value. At $\Lambda=750~\mathrm{MeV}$ with NR $\pi$, the calibration gives $\lambda_{0,\Lambda}=3.21892~\mathrm{GeV}^{-1}$ and $\Gamma_{D\pi}^{0}=35.4~\mathrm{keV}$. The corresponding bare masses are $m_{D^*,0,\Lambda}^{+}=2019.071~\mathrm{MeV}$ and $m_{D^*,0,\Lambda}^{0}=2015.512~\mathrm{MeV}$.

\subsection{Two-body interactions and pion time orderings}\label{sec:timeorderings}

The calibrated $D^*D\pi$ vertex also fixes the OPE interaction between the two heavy mesons. As shown in Fig.~\ref{fig:timeorderings} of the Letter, the OPE interaction has two time orderings. For the static matching, the transferred pion energy is $q_0=\Delta$. The denominators of the two orderings obey the identity
\begin{equation}
\frac{1}{\Delta^2-\omega_Q^2}
=\frac{1}{2\omega_Q}
\left[\frac{1}{\Delta-\omega_Q}-\frac{1}{\Delta+\omega_Q}\right],
\label{eq:timeorderingidentity}
\end{equation}
where $\bm Q$ is the transferred spatial momentum and $\omega_Q=\sqrt{m_\pi^2+Q^2}$. In the time ordering A, the initial $D^*$ emits the pion first and the intermediate state is $DD\pi$. Its propagation is already generated by $W_{32}$ and the explicit three-body block. In the time ordering B, the intermediate state is $D^*D^*\pi$, and its high-energy denominator is represented by the static term $V_B^S$ in $H_{22}$ [Eq.~\eqref{eq:h22} of the Letter]. This assignment includes each pion time ordering once. With the $f_\Lambda(Q)^2$ regulator, the momentum-space static B potential is
\begin{equation}
V_{B,I}^{S}(Q)=-\frac{\xi_I\lambda_{0,\Lambda}^2}{3}
\frac{Q^2e^{-2Q^2/\Lambda^2}}{2\omega_Q(\omega_Q+\Delta)}.
\label{eq:staticb}
\end{equation}
The exchange coefficients are $(\xi_0,\xi_1)=(3,1)$ for $DD^*$ and $(3,-1)$ for the $C=+$ $D\bar D^*$ channels in our field convention. For the heavy-meson separation $r=|\bm R|$, the coordinate-space form reads
\begin{align}
Y_{B,\Lambda}(r)&=\frac{1}{2\pi^2}\int_0^\infty\dd Q\,Q^2j_0(Qr)
\frac{e^{-2Q^2/\Lambda^2}}{2\omega_Q(\omega_Q+\Delta)},\nonumber\\
V_{B,I}^{S}(r)&=\frac{\xi_I\lambda_{0,\Lambda}^2}{3}
\left[Y_{B,\Lambda}''(r)+\frac{2}{r}Y_{B,\Lambda}'(r)\right],
\label{eq:coordinateb}
\end{align}
where $j_0$ is the spherical Bessel function.

The regulated contact interaction in the isospin channel $I$ reads
\begin{align}
V_{C,I}(Q)&=C_Ie^{-2Q^2/\Lambda^2},\nonumber\\
V_{C,I}(r)&=C_I\frac{\Lambda^3}{16\sqrt2\pi^{3/2}}
e^{-\Lambda^2r^2/8}
\label{eq:contactpotential}
\end{align}
in the momentum and coordinate space, respectively. Section~\ref{app:charge} gives the charge-basis contact matrices.

\subsection{Contact calibration}\label{sec:chargechannels}

We take $C_1=C_1^X=0$ as a baseline. The isoscalar contacts $C_0$ and $C_0^X$ are fitted independently in the $\Tcc$ and $\Xstate$ systems, respectively. We use the pole convention
\begin{equation}
\epole=M-\frac{i}{2}\Gamma,\qquad
\Gamma=-2\operatorname{Im}(\epole),
\label{eq:poleconvention}
\end{equation}
where $M$ is the pole mass and $\Gamma$ is the width. For $\Tcc$, the experimental pole mass offset~\cite{LHCb:2021auc} and the masses in Table~\ref{tab:masses} give the target $3874.750~\mathrm{MeV}$. The central value of the measured $\Xstate$ pole mass~\cite{BESIII:2023hml} lies almost exactly at the neutral $D^0\bar D^{*0}$ threshold. We therefore search for pole masses at both $3871.60$ and $3871.80~\mathrm{MeV}$, which are $100~\mathrm{keV}$ below and above that threshold, respectively. For each cutoff and pion kinetic scheme, we fit the real $C_0$ and $C_0^X$ to the respective mass targets. The widths are predicted. The results are presented in Sec.~\ref{sec:poles}.

\section{Gaussian expansion and complex scaling}\label{sec:methods}

\subsection{Gaussian expansion method}\label{sec:gem}

\begin{figure}[htbp]
\centering
\includegraphics[width=0.5\columnwidth]{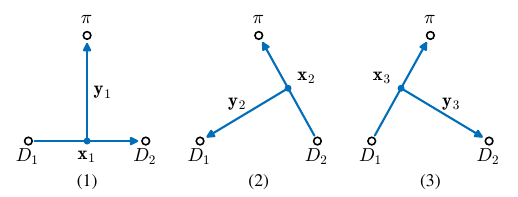}
\caption{The three Jacobi rearrangements of $D_1D_2\pi$. The pair is described by its internal coordinate $\bm x_c$, and the spectator is connected to the center of mass of the remaining pair by $\bm y_c$. The same construction applies to $D\bar D\pi$.}
\label{fig:jacobi}
\end{figure}

We expand both Fock components in Gaussian basis functions~\cite{Hiyama:2003cu}, including all three Jacobi rearrangements of Fig.~\ref{fig:jacobi} for the three-body channels. For a pair $(i,j)$ with spectator $k$, the coordinates are
\begin{equation}
\bm x_c=\bm r_j-\bm r_i,\qquad
\bm y_c=\bm r_k-\frac{m_i\bm r_i+m_j\bm r_j}{m_i+m_j},
\label{eq:jacobidefinition}
\end{equation}
and the rearrangements are $DD$--$\pi$, $D_2\pi$--$D_1$, and $D_1\pi$--$D_2$. For $\Xstate$, one heavy meson is replaced by its antiparticle.

The normalized single coordinate basis functions read
\begin{align}
\phi^G_{nlm}(\bm r)=N_{nl}r^le^{-\nu_nr^2}Y_{lm}(\hat{\bm r}),\qquad
N_{nl}=\left[\frac{2(2\nu_n)^{l+3/2}}{\Gamma(l+3/2)}\right]^{1/2},
\qquad \nu_n=r_n^{-2},
\label{eq:gaussianbasis}
\end{align}
with the ranges in geometric progression,
\begin{equation}
r_n=r_{\min}a^{n-1},\qquad
a=\left(\frac{r_{\max}}{r_{\min}}\right)^{1/(n_{\max}-1)}.
\label{eq:gaussianranges}
\end{equation}
For every relative coordinate we use $n_{\max}=50$ functions with $r_{\min}=0.02~\mathrm{fm}$ and $r_{\max}=50~\mathrm{fm}$. The same basis configuration applies to all cutoffs and both pion kinetic schemes.

The two-body and three-body components are expanded as
\begin{align}
\Psi_2&=\mathcal S\sum_{\alpha,n}a_{\alpha n}
\big[\phi^G_{n0}(\bm R)\otimes\chi_{s=1}\big]_{J=1}|\alpha\rangle,
\label{eq:twowavefunction}
\\
\Psi_3&=i\mathcal S\sum_{\gamma,c,n,m}\sum_{(l_x,l_y)}b_{\gamma cnm}^{l_xl_y}
\big[\phi^G_{nl_x}(\bm x_c)\otimes
\phi^G_{ml_y}(\bm y_c)\big]_{J=1}|\gamma\rangle,
\label{eq:threewavefunction}
\end{align}
respectively. Here $\alpha$ and $\gamma$ label the two- and three-body charge channels, $c$ labels the Jacobi rearrangement, and $n,m$ label the Gaussian ranges. The $\chi_{s=1}$ is the vector-meson spin wave function, and $J$ is the total angular momentum. The factor $i$ in three-body components is a phase factor that makes the transition matrices real. The three-body components contain $(l_x,l_y)=(1,0)$ and $(0,1)$ modes in each rearrangement, both giving total $L=1$.

$\mathcal{S}$ is the symmetrization operator corresponding to each component. For $\Tcc$, $\mathcal S=(1+P_{12})/2$ symmetrizes the identical $D$ mesons in the $D^0D^0\pi^+$ and $D^+D^+\pi^-$ channels. The $D^0$ and $D^+$ in the $D^0D^+\pi^0$ channel are distinguishable, and no exchange symmetry is imposed on their spatial wave function. For $\Xstate$, $\mathcal S=(1+C)/2$ projects the charge and coordinate wave function onto $C=+$. The two charged-pion channels form a single $C=+$ component:
\begin{align}
\Psi_{\pi^\pm}=\frac{1}{\sqrt2}\big[\psi(\bm r_1,\bm r_2,\bm r_\pi)|D^0D^-\pi^+\rangle+\psi(\bm r_2,\bm r_1,\bm r_\pi)|D^+\bar D^0\pi^-\rangle\big].
\label{eq:chargedcprojection}
\end{align}
In the $D^0\bar D^0\pi^0$ and $D^+D^-\pi^0$ channels, this requires the spatial wave function to be symmetric under exchange of the meson and antimeson.

Each spatially symmetric three-body block contains $3\times50^2=7500$ basis functions. The $D^0D^+\pi^0$ component of $\Tcc$ and the charged-pion $C=+$ component of $\Xstate$ each contain $6\times50^2=15000$ basis functions. Each two-body block contains $50$ basis functions, giving a total of $30100$ basis functions for each system before overlap reduction. We discard overlap eigenvectors whose eigenvalues are smaller than $10^{-10}$. This removes zero-norm and linearly dependent directions, leaving $94$ two-body basis states and $19397$ ($19402$) three-body basis states for $\Tcc$ ($\Xstate$). The resulting matrix dimensions are $19491$ and $19496$, respectively.

We also tested $n_{\max}=60$, $r_{\min}=0.01~\mathrm{fm}$, and $r_{\max}=60~\mathrm{fm}$ separately. For both $\Tcc$ and $\Xstate$, the pole positions shifted by less than $1~\mathrm{keV}$. These checks confirm numerical convergence, and the residual numerical uncertainties are negligible at the precision of the present study.

\subsection{Complex scaling method}\label{sec:csm}

In a finite basis expansion, the continuum above an open threshold also appears as discrete eigenvalues. A scheme is required to distinguish the genuine discrete poles from such discretized continuum states. Moreover, the resonance wave functions are not square integrable in the original coordinates. We use CSM~\cite{Aguilar:1971ve,Balslev:1971vb,Moiseyev:1998gjp,Aoyama:2006hrz,Chen:2023eri} to address both points. This technique introduces a complex rotation of the radial coordinate and its conjugate momentum,
\begin{equation}
U(\theta)r=re^{i\theta},\qquad
U(\theta)p=pe^{-i\theta},
\label{eq:coordinate-momentum-rotation}
\end{equation}
where $\theta$ is the scaling angle. For an $N$-body wave function with $N-1$ Jacobi coordinates, the spatial dimension is $d_N=3(N-1)$. Under complex scaling, the wave function transforms as
\begin{equation}
[U_N(\theta)\psi_N](\bm\xi_N)
=e^{id_N\theta/2}\psi_N(e^{i\theta}\bm\xi_N),
\label{eq:dilation}
\end{equation}
where $\bm\xi_N$ collects all Jacobi coordinates. The factor $e^{id_N\theta/2}$ compensates for the change in the integration measure.

The full transformation is $U(\theta)=\operatorname{diag}(U_2,U_3)$, and the blocks transform as
\begin{equation}
H_{NN}^{\theta}=U_NH_{NN}U_N^{-1},\qquad
W_{32}^{\theta}=U_3W_{32}U_2^{-1}.
\label{eq:blockscaling}
\end{equation}
The inverse transition is transformed with $U_2W_{23}U_3^{-1}$ and $W_{23}^\theta=(W_{32}^\theta)^T$. The different spatial dimensions of the two sectors enter the scaling of the transition block. For the relativistic pion, the kinetic operator becomes
\begin{equation}
T_\pi^{\Rpi,\theta}(p)=\sqrt{m_\pi^2+e^{-2i\theta}p^2}-m_\pi.
\label{eq:relativescaling}
\end{equation}

After complex scaling and the expansion in the Gaussian basis, the coupled Schr\"odinger equation reads
\begin{equation}
\begin{pmatrix}H_{22}^\theta&W_{23}^\theta\\W_{32}^\theta&H_{33}^\theta\end{pmatrix}
\begin{pmatrix}\bm a\\\bm b\end{pmatrix}
=E\begin{pmatrix}N_2&0\\0&N_3\end{pmatrix}
\begin{pmatrix}\bm a\\\bm b\end{pmatrix}.
\label{eq:gemeigenproblem}
\end{equation}
Here $\bm a$ and $\bm b$ collect the expansion coefficients, and $N_2$ and $N_3$ are the overlap matrices of the two sectors. Since no matrix element depends on $E$, \eq{eq:gemeigenproblem} is a linear generalized eigenvalue problem. The poles and discretized scattering states can be obtained by a single diagonalization.

\begin{figure}[t]
\centering
\includegraphics[width=0.5\columnwidth]{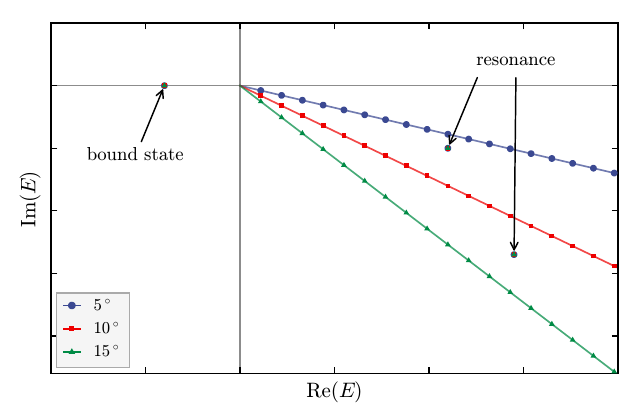}
\caption{A typical solution of the complex-scaled Schr\"odinger equation. The discretized scattering eigenvalues follow the continuum rays rotated by $-2\theta$, while the bound state and resonance eigenvalues remain at their stationary positions.}
\label{fig:csm-typical}
\end{figure}

Under complex scaling, the discretized continuum states lie along a $\theta$-dependent line starting at the threshold. In the NR scheme, this line is given by $E_{\mathrm{th}}+e^{-2i\theta}t$, with $t\geq0$. The three-body thresholds are sums of the constituent particle masses. The continuum states associated with $D^*$ start at $m_D+z_*$, where $z_*$ denotes the physical $D^*$ pole determined in Sec.~\ref{sec:vertex-calibration}. These starting points are complex. The pole of interest remains stationary as the scaling angle varies, while the discretized continua rotate, as schematically illustrated in Fig.~\ref{fig:csm-typical}.

\section{\texorpdfstring{Benchmark details for $T_{cc}^{+}$ and $X(3872)$}{Benchmark details for Tcc+ and X(3872)}}\label{sec:results}

\subsection{Pole spectra and parameter dependence}\label{sec:poles}

We set $\Lambda=750~\mathrm{MeV}$, NR $\pi$, and $C_1=C_1^X=0$ as the default configuration. For $\Tcc$ and $\Xstate$, we use $\theta=5^\circ$, $10^\circ$, and $15^\circ$ for the pole identification. The identified poles are stable at the $0.1~\mathrm{keV}$ scale across the three angles. We compare the calculated complex eigenvalues with the experimental $T$-matrix pole positions~\cite{LHCb:2021auc,BESIII:2023hml}, rather than with the Breit--Wigner parameters.

The cutoff scan in Fig.~\ref{fig:lambda} of the Letter covers $600$ to $900~\mathrm{MeV}$ with spacing $25~\mathrm{MeV}$. At each cutoff, the $D^*$ bare parameters are recalibrated to their physical inputs and the real isoscalar contact is refitted to the pole masses. The $\Tcc$ width changes from $55$ to $51~\mathrm{keV}$ for NR $\pi$ and from $54$ to $50~\mathrm{keV}$ for R $\pi$. The corresponding $\Xstate$ widths change from $38$ to $37~\mathrm{keV}$ for both pion kinetic schemes. Thus the results are essentially independent of the cutoff, and the treatment of the pion kinetic energy has little effect on the results.

\begin{figure}[t]
\centering
\includegraphics[width=0.8\columnwidth]{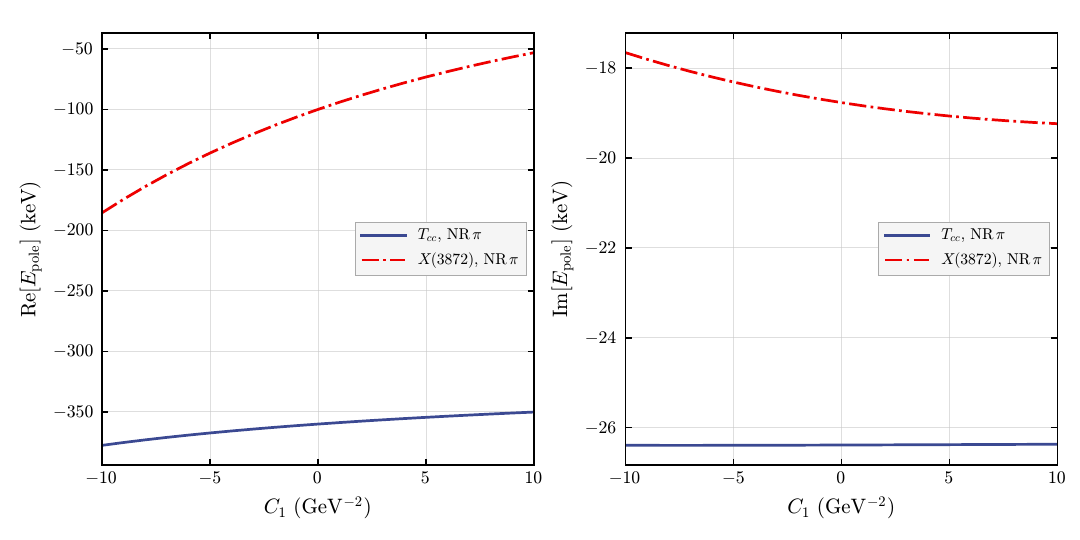}
\caption{Isovector contact $C_1^{(X)}$ dependence of the poles at $\Lambda=750~\mathrm{MeV}$, NR $\pi$, and $\theta=10^\circ$. The isoscalar contacts in Table~\ref{tab:poles750nr} of the Letter are fixed during the scan. The left panel gives the pole real part offsets to the $D^0D^{*+}$ and $D^0\bar{D}^{*0}$ thresholds for $\Tcc$ and $\Xstate$, respectively. The right panel gives the pole imaginary part.}
\label{fig:c1}
\end{figure}

Figure~\ref{fig:c1} shows the isovector contact scan at $\Lambda=750~\mathrm{MeV}$ with NR $\pi$. We vary $C_1$ and $C_1^X$ from $-10$ to $10~\mathrm{GeV}^{-2}$ with the $C_0$ and $C_0^X$ in Table~\ref{tab:poles750nr} of the Letter fixed. For $\Tcc$, the pole mass shifts from $-18$ to $+10~\mathrm{keV}$, and the width changes by less than $0.1\%$. For $\Xstate$, the mass shifts from $-85$ to $+47~\mathrm{keV}$, and the width shifts from $-5.9\%$ to $+2.5\%$. Although $\Xstate$ is more sensitive to its isovector contact than $\Tcc$, it remains at the scale of tens of keV for the mass and at the percent level for the width throughout the scan. The results support the adoption of $C_1^{(X)}=0$ as a suitable baseline.

\subsection{Channel compositions and spatial structures}\label{sec:structure}

\begin{table*}[t]
\caption{Channel compositions and pairwise rms distances at $\Lambda=750~\mathrm{MeV}$, NR $\pi$, $C_1=C_1^X=0$, and $\theta=10^\circ$. $R_{12}$ is the heavy-meson separation, and $R_{1\pi}$ and $R_{2\pi}$ follow the order of the two heavy mesons in each row. For the two-body channels, $R_{12}$ is the pseudoscalar--vector separation.}
\label{tab:structure}
\begin{tabular*}{\textwidth}{@{\extracolsep{\fill}}llrrrr@{}}
\hline\hline
System & Channel & $P_\alpha$ (\%) & $R_{12}$ (fm) & $R_{1\pi}$ (fm) & $R_{2\pi}$ (fm)\\
\hline
$\Tcc$ & $D^0D^{*+}$ & 70.9 & 5.70 & --- & --- \\
 & $D^+D^{*0}$ & 25.9 & 2.80 & --- & --- \\
 & $D^0D^0\pi^+$ & 1.7 & 5.21 & 4.39 & 4.39 \\
 & $D^0D^+\pi^0$ & 1.1 & 4.76 & 4.82 & 2.68 \\
 & $D^+D^+\pi^-$ & 0.5 & 2.80 & 3.40 & 3.40 \\
\hline
$\Xstate$ & $D^0\bar D^{*0},\ D^{*0}\bar D^0$ & 91.4 & 10.37 & --- & --- \\
 & $D^+D^{*-},\ D^{*+}D^-$ & 5.9 & 1.53 & --- & --- \\
 & $D^0\bar D^0\pi^0$ & 0.9 & 9.68 & 5.77 & 5.77 \\
 & $D^+D^-\pi^0$ & 0.1 & 1.68 & 2.79 & 2.79 \\
 & $D^0D^-\pi^+$ & 0.9 & 9.93 & 10.38 & 3.30 \\
 & $D^+\bar D^0\pi^-$ & 0.9 & 9.93 & 3.30 & 10.38 \\
\hline\hline
\end{tabular*}
\end{table*}

\begin{figure*}[t]
\centering
\includegraphics[width=\textwidth]{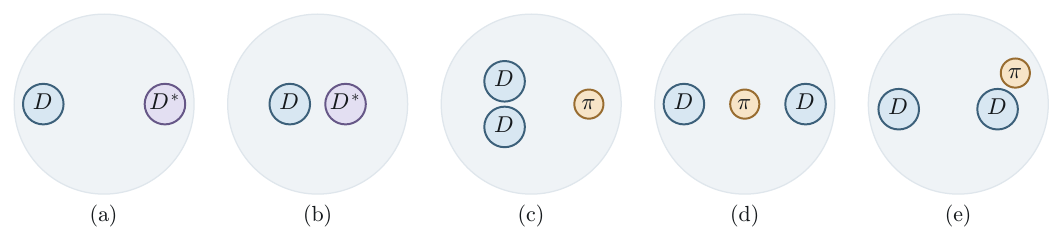}
\caption{Schematic spatial configurations of the wave function: (a) an extended pseudoscalar--vector component; (b) a localized pseudoscalar--vector component; (c) a close heavy-meson pair with an outer pion; (d) two separated heavy mesons sharing the pion; (e) a correlated $D\pi$ subcluster with a distant $D$ spectator. For $\Xstate$, one heavy meson should be understood as the corresponding antimeson.}
\label{fig:molecular}
\end{figure*}

The coupled eigenvector gives the wave function in every charge channel. For the complex-scaled pole, we use the $c$-product, which is bilinear in the expansion coefficients and does not conjugate them~\cite{Moiseyev:1998gjp}. For charge channel $\alpha$ in either Fock sector, let $\bm c_\alpha$ be its Gaussian coefficient vector and $N_\alpha$ its overlap matrix. The total normalization and channel compositions are
\begin{align}
\mathcal N&=\sum_\alpha\bm c_\alpha^TN_\alpha\bm c_\alpha,\nonumber\\
P_\alpha&=\frac{\bm c_\alpha^TN_\alpha\bm c_\alpha}{\mathcal N}.
\label{eq:channelweights}
\end{align}
The sum runs over two- and three-body charge channels and satisfies $\sum_\alpha P_\alpha=1$.

The pairwise rms distance within channel $\alpha$ is
\begin{equation}
R_{ij,\alpha}=\left[
\frac{\bm c_\alpha^T\mathcal R_{ij,\alpha}^{\theta}\bm c_\alpha}
{\bm c_\alpha^TN_\alpha\bm c_\alpha}
\right]^{1/2},
\label{eq:radii}
\end{equation}
where the radius matrix is
\begin{equation}
\big[\mathcal R_{ij,\alpha}^{\theta}\big]_{uv}
=e^{2i\theta}\langle\phi_{\alpha u}|r_{ij}^2|\phi_{\alpha v}\rangle.
\label{eq:radiusmatrix}
\end{equation}
Here $r_{ij}=|\bm r_i-\bm r_j|$, and $u,v$ label the basis functions in channel $\alpha$. Each radius is normalized to its own channel, while $P_\alpha$ is normalized to the complete Fock state. The factor $e^{2i\theta}$ is the CSM transformation of $r_{ij}^2$. For resonance states these quantities can be complex. We list the real parts of $P_\alpha$ and $R_{ij,\alpha}$ in Table~\ref{tab:structure} for the qualitative discussion.

As listed in Table~\ref{tab:structure}, the two-body sector carries $96.7\%$ of the $\Tcc$ weight and $97.3\%$ of the $\Xstate$ weight. The two $\Tcc$ two-body channels have the radii $5.70$ and $2.80~\mathrm{fm}$. The neutral and charged $\Xstate$ channels have $10.37$ and $1.53~\mathrm{fm}$. The separation of scales follows the offsets to the respective charge thresholds. The $\Xstate$ pole lies $0.1~\mathrm{MeV}$ below the neutral threshold and $8.33~\mathrm{MeV}$ below the charged one. The offsets for $\Tcc$ are $0.36$ and $1.77~\mathrm{MeV}$. The lower threshold component is spatially extended, and the higher threshold component is localized. Similar differences between the component radii of $\Tcc$ and $\Xstate$ appear in coupled-channel calculations~\cite{Takeuchi:2014rsa,Cheng:2022qcm}.

The explicit three-body component contributes $3.3\%$ for $\Tcc$ and $2.7\%$ for $\Xstate$. Their channel radii display three spatial structures, illustrated schematically in Fig.~\ref{fig:molecular}. The structures that contain a close heavy-meson pair with an outer pion are realized in the $D^+D^+\pi^-$ and $D^+D^-\pi^0$ channels. The structures with two well-separated heavy mesons sharing one pion are realized in the $D^0D^0\pi^+$, $D^0D^+\pi^0$ and $D^0\bar D^0\pi^0$ channels. A correlated $D\pi$ subcluster with a distant $D$ spectator is realized in the other channels.

\section{Bottom-sector details}\label{sec:bottom}

For the bottom systems, we work in the isospin limit with NR $\pi$. At each cutoff, we inherit the contact coefficients from the $\Tcc$ and $\Xstate$ fits, as given in Eq.~\eqref{eq:bottomcontacts} of the Letter. For the $(DB^*,D^*B)$ and $(D\bar B^*,D^*\bar B)$ systems, we neglect the off-diagonal contact interaction, since the thresholds of the two channels differ by $96.13~\mathrm{MeV}$. Their off-diagonal interaction is generated by pion exchange. The time ordering A proceeds through the explicit three-body space, and the static time ordering B also contributes to the off-diagonal two-body block. For the latter, we use the average of the two directional static denominators
\begin{equation}
\frac{1}{2\omega_Q}\frac{1}{2}\left[\frac{1}{\omega_Q+\Delta_D}+\frac{1}{\omega_Q+\Delta_B}\right].
\label{eq:averagedenominator}
\end{equation}

The isospin masses of $D$, $D^*$, and $\pi$ are those of \eq{eq:isomasses}. We use $m_B=5279.565~\mathrm{MeV}$ and $m_{B^*}=5324.75~\mathrm{MeV}$ from the PDG central values~\cite{ParticleDataGroup:2026mpi}. The $D^*$ bare parameters reproduce its isospin-limit mass and width. For the $B^*$ axial coupling, we use the leading heavy-flavor relation $g_{B,a}=g_a$ for the physical value~\cite{Neubert:1993mb}. The physical coupling $\lambda_a=g_a/(2f_\pi)$ is given in \eq{eq:physicalcoupling}. We fix the bare mass $m_{B^*,0,\Lambda}$ and the bare coupling $\lambda_{B,0,\Lambda}$ by matching the physical mass and the pole residue,
\begin{equation}
\begin{aligned}
m_{B^*}&=m_{B^*,0,\Lambda}+\Sigma_B(m_{B^*}),\\
\sqrt{Z_B}\lambda_{B,0,\Lambda}&=\lambda_a,\qquad
Z_B=\left[1-\Sigma_B'(m_{B^*})\right]^{-1}.
\end{aligned}
\label{eq:bstarcalibration}
\end{equation}
Here $\Sigma_B$ is the self-energy of \eq{eq:starselfenergy} with the daughter $D$ replaced by $B$, and the prime denotes differentiation with respect to the subsystem energy. The $B^*$ lies $92.854~\mathrm{MeV}$ below $B\pi$, so its self-energy and bare parameters are real. Its strong decay width vanishes. At $\Lambda=750~\mathrm{MeV}$, we obtain $m_{B^*,0,\Lambda}=5331.706~\mathrm{MeV}$, $\lambda_{B,0,\Lambda}=3.08672~\mathrm{GeV}^{-1}$ and $Z_B=0.9902$. The two-body kinetic terms use the corresponding bare vector masses $m_{D^*,0,\Lambda}$ and $m_{B^*,0,\Lambda}$. The same bare vertices enter constituent dressing, two--three-body transitions, and the static pion exchange of time ordering B.


We use the same $50$ Gaussian functions over $0.02$--$50~\mathrm{fm}$ for each coordinate. We impose Bose symmetrization for the $bb$ system and $C=+$ projection for the $b\bar{b}$ system. The $bb(\bar{b})$ systems contain $7550$ basis functions and the $cb(\bar{b})$ systems contain $15100$ basis functions. We use $\theta=20^\circ$ and $25^\circ$ for the pole identification.

A bound state appears below the lowest two-body threshold in each system, as shown in Fig.~\ref{fig:prl-spectra} of the Letter. We list the four lowest states in Table~\ref{tab:bottompoles}.

\begin{table}[htbp]
\centering
\caption{Bound states at $\Lambda=750~\mathrm{MeV}$. $M$ is the mass and $E_B$ is the binding energy from the lowest two-body threshold.}
\label{tab:bottompoles}
\begin{tabular*}{\columnwidth}{@{\extracolsep{\fill}}lrr@{}}
\hline\hline
System & $M$ (MeV) & $E_B$ (MeV)\\
\hline
$c\bar b$ & $7187.153$ & $-4.847$\\
$cb$ & $7189.296$ & $-2.704$\\
$bb$ & $10579.768$ & $-24.547$\\
$b\bar b$ & $10574.669$ & $-29.646$\\
\hline\hline
\end{tabular*}
\end{table}

The cutoff dependence of these poles is shown in Fig.~\ref{fig:prl-cutoff} of the Letter. The two resonance masses each vary by about $1~\mathrm{MeV}$ across the cutoff range. Their imaginary parts are more cutoff-sensitive, corresponding to widths of $0.183$--$1.685~\mathrm{MeV}$ and $0.151$--$1.279~\mathrm{MeV}$ for the $c\bar b$ and $cb$ resonances, respectively. The placement of the two poles on opposite sides of the three-body threshold is preserved throughout.

Using the same $c$-product definitions as in Eqs.~\eqref{eq:channelweights}--\eqref{eq:radiusmatrix}, we evaluate the channel compositions and pairwise rms distances for all six poles at $\theta=20^\circ$. Table~\ref{tab:bottomstructure} lists their real parts for a qualitative illustration. The $cb(\bar{b})$ bound states are dominated by $DB^*$ and $D\bar B^*$, whereas the resonances are dominated by $D^*B$ and $D^*\bar B$. Their dominant channel separations are about $1.6$--$2.0~\mathrm{fm}$. The $bb(\bar{b})$ bound states are more localized with $0.74$--$0.78~\mathrm{fm}$ spatial distances. The explicit three-body weights are $0.9$--$2.0\%$. In the $BB(\bar{B})\pi$ three-body components, the pion extends farther from each heavy meson than the heavy-pair separation. In the $cb(\bar{b})$ bound states, the pion--bottom rms separation is smaller than the pion--charm separation. The resonances reverse this ordering and have larger pion--heavy-meson distances.

\begin{table*}[t]
\centering
\setlength{\tabcolsep}{4pt}
\caption{Channel compositions and pairwise rms distances of the four bound states and two resonances at $\Lambda=750~\mathrm{MeV}$. $R_{12}$ is the heavy-meson separation, and $R_{1\pi},R_{2\pi}$ follow the heavy-meson order in each three-body channel.}
\label{tab:bottomstructure}
\begin{tabular*}{\textwidth}{@{\extracolsep{\fill}}llrrrr@{}}
\hline\hline
State & Channel & $P_\alpha$ (\%) & $R_{12}$ (fm) & $R_{1\pi}$ (fm) & $R_{2\pi}$ (fm)\\
\hline
$c\bar b$ (bound) & $DB^*$ & 98.7 & 1.61 & --- & --- \\
 & $D^*B$ & 0.3 & 0.68 & --- & --- \\
 & $DB\pi$ & 0.9 & 1.64 & 2.13 & 1.41 \\
\hline
$cb$ (bound) & $D\bar B^*$ & 98.8 & 2.01 & --- & --- \\
 & $D^*\bar B$ & 0.2 & 0.69 & --- & --- \\
 & $D\bar B\pi$ & 0.9 & 2.04 & 2.47 & 1.43 \\
\hline
$bb$ (bound) & $BB^*$ & 98.7 & 0.78 & --- & --- \\
 & $BB\pi$ & 1.3 & 0.77 & 1.47 & 1.47 \\
\hline
$b\bar b$ (bound) & $B\bar B^*$ & 98.7 & 0.74 & --- & --- \\
 & $B\bar B\pi$ & 1.3 & 0.73 & 1.44 & 1.44 \\
\hline
$c\bar b$ (resonance) & $DB^*$ & 0.2 & 0.47 & --- & --- \\
 & $D^*B$ & 98.0 & 1.59 & --- & --- \\
 & $DB\pi$ & 1.7 & 1.88 & 2.73 & 3.15 \\
\hline
$cb$ (resonance) & $D\bar B^*$ & 0.2 & 0.48 & --- & --- \\
 & $D^*\bar B$ & 97.8 & 2.00 & --- & --- \\
 & $D\bar B\pi$ & 2.0 & 2.42 & 3.89 & 4.31 \\
\hline\hline
\end{tabular*}
\end{table*}

\section{Charge-channel transition coefficients}\label{app:charge}

The particle vertices extracted from \eq{eq:vertex} carry the charge coefficients
\begin{align}
D^{*0}\to(D^0\pi^0,D^+\pi^-):(1,\sqrt2),\qquad
D^{*+}\to(D^0\pi^+,D^+\pi^0):(\sqrt2,-1),
\label{eq:chargevertices}
\end{align}
and the antiparticle vertices follow from the field phases used in \eq{eq:cchannels}. Let $w_i$ denote a single matrix element with daughter $D_i$, including $\lambda_{0,\Lambda}$ and the regulator but without the charge and symmetry coefficients. For the symmetrized $\Tcc$ basis, with the two-body columns $(D^0D^{*+},D^+D^{*0})$ and the three-body rows $(D^0D^0\pi^+,D^0D^+\pi^0,D^+D^+\pi^-)$, the transition reads
\begin{equation}
W_{32}^{T}=\begin{pmatrix}
2w_2&0\\-w_2&w_1\\0&2w_2
\end{pmatrix}.
\label{eq:wtcc}
\end{equation}
For $\Xstate$, the columns are the $(n,c)$ states of \eq{eq:cchannels}, and the rows are $(D^0\bar D^0\pi^0,D^+D^-\pi^0,\pi^\pm\text{ pair})$, where the last row uses the joint $C=+$ state of \eq{eq:chargedcprojection}. The transition reads
\begin{equation}
W_{32}^{X}=\begin{pmatrix}
-\sqrt2w_2&0\\0&+\sqrt2w_2\\-\sqrt2w_2&-\sqrt2w_1
\end{pmatrix}.
\label{eq:wx}
\end{equation}

For the symmetrized rows, the two daughter-pair matrix elements are equal, while the two paths in a mixed-charge row remain distinct and enter coherently. The factors $2$ and $\sqrt2$ arise from the different Bose and $C$ embeddings.

The charge-basis contact matrices in these conventions read
\begin{align}
V_C^{T}(Q)=\frac{e^{-2Q^2/\Lambda^2}}2
\begin{pmatrix}C_0+C_1&C_1-C_0\\C_1-C_0&C_0+C_1\end{pmatrix},\qquad
V_C^{X}(Q)=\frac{e^{-2Q^2/\Lambda^2}}2
\begin{pmatrix}C_0^X+C_1^X&C_0^X-C_1^X\\C_0^X-C_1^X&C_0^X+C_1^X\end{pmatrix}.
\label{eq:chargecontacts}
\end{align}
Thus $C_I$ and $C_I^X$ are the normalized isospin eigenvalues of their respective systems, while the charge-mass splittings mix the wave-function components through propagation.

\end{document}